\documentclass{svjour3}                     
\usepackage{graphicx}
\usepackage{rotating}
\usepackage{amssymb}
\usepackage{amsmath}
\usepackage{mathptmx}
\usepackage[numbers]{natbib}
\usepackage{bm}
 \journalname{Journal of Low Temperature Physics}
\begin{document}

\title{Pseudogap phenomena near the BKT transition of a two-dimensional
ultracold Fermi gas in the crossover region
}


\author{M. Matsumoto\and R. Hanai\and D. Inotani\and Y. Ohashi}


\institute{\at
Faculty of Science and Technology, Keio University, 3-14-1 Hiyoshi, Kohoku-ku, Yokohama 223-8522, Japan. \\
Tel.: +81-45-566-1454 \\
Fax: +81-45-566-1672 \\
\email{moriom@rk.phys.keio.ac.jp}
}

\date{Received: \today / Acceptted: \today}

\maketitle
\begin{abstract}
We investigate strong-coupling properties of a two-dimensional ultracold
 Fermi gas in the normal phase.
 In the  three-dimensional case, it has been shown that 
 the so-called pseudogap phenomena can be well described by a 
 (non-self-consistent) $T$-matrix approximation (TMA).
 In the two-dimensional case,
while this strong coupling theory
 can explain the pseudogap phenomenon in the strong-coupling
 regime, it unphysically gives large pseudogap size 
 in the crossover region, as well as in the weak-coupling regime.
 We show that this difficulty can be overcome when one improve TMA to include
 higher order pairing fluctuations within the framework of a
 self-consistent $T$-matrix approximation (SCTMA).
 The essence of this improvement is also explained.
 Since the observation of the BKT transition has recently been reported
 in a two-dimensional $^6$Li Fermi gas, our results would be useful for
 the study of strong-coupling physics  associated with this
 quasi-long-range order.
\end{abstract}
\par
\section{Introduction}
\par
An ultracold Fermi gas is well known as a system with high
tunability of various physical parameters\cite{Gurarie,Bloch}.
For example, one can experimentally tune the strength of a pairing interaction associated with a
Feshbach resonance.
This has enabled us to systematically study how superfluid properties
continuously change from the weak-coupling BCS (Bardeen-Cooper-Schrieffer
)-type to the BEC (Bose-Einstein condensation) of tightly bound
molecular boson with increasing the interaction strength, which
is also referred to as the BCS-BEC crossover in the literature.
In the crossover region, pairing fluctuations are expected to be strong
near the superfluid phase transition temperature $T_{\rm c}$, so that the
so-called pseudogap phenomenon has been discussed there\cite{Stewart,Gaebler}.
\par
Another example of the high tunability is the realization of a low
dimensional Fermi gas by using an optical lattice technique.
Since pairing fluctuations are enhanced by the low dimensionality of the
system, together with the tunable pairing interaction, this is also use
for the study of strong-coupling physics in a systematic manner.
In particular, two-dimensional Fermi gases have recently attracted much
attention in this
field\cite{Martiyanov,Feld,Frohlich,Sommer,Makhalov,Ries,Murthy2,Fenech},
because the quasi-long-range superfluid order, called the
Berezinskii-Kosterlitz-Thouless (BKT) phase is expected there\cite{Berezinskii,Kosterlitz}.
Indeed, various physical quantities, such as photoemission
spectra\cite{Feld} , as well as thermodynamic
quantities\cite{Makhalov,Fenech}, have been measured in this system, and
the observation of BKT transition has recently been reported in a
two-dimensional $^6$Li Fermi gas\cite{Ries,Murthy2}.
\par
In the three-dimensional case, a (non-self-consistent) $T$-matrix
approximation (the detail of which is explained in Sec. 2) has
been
extensively
used to successfully explain various interesting phenomena
observed in the BCS-BEC crossover region \cite{Tsuchiya,Chen,HuiHu}. 
In this regard, however, when TMA is applied to the two-dimensional
case, while it is valid for the strong-coupling regime, it overestimates
strong-coupling effects in the weak-coupling
regime\cite{Marsiglio,Matsumoto1}. 
For example, TMA does not give free-particle density of states even in
the weak-coupling regime, when the pairing interaction is very weak.
Thus, in the current stage of research, it is a crucial theoretical
issue to improve TMA so that one can correctly deal with the
weak-coupling regime.
\par
In this paper, we show that the self-consistent $T$-matrix approximation
(SCTMA), which involves higher-order pairing fluctuations than TMA,
meets our demand.
Within this framework, we clarify how the so-called pseudogap phenomenon
disappears in a two-dimensional Fermi gas as one approaches the
weak-coupling regime from the strong-coupling side.
Comparing SCTMA results with TMA ones, we also discuss the reason why
the above mentioned problem in TMA can be eliminated in SCTMA.
\par
Throughout this paper, we take $\hbar =k_{\rm B}=1$ and the two-dimensional system area is taken to be unity, for simplicity.
\par
\section{Formulation}
\par
We consider a two-dimensional uniform Fermi atomic gas consisting of two atomic hyperfine states, described by the BCS Hamiltonian,
\begin{align}
 H=\sum_{\bm{p},\sigma}\xi_{\bm{p}}c^{\dagger}_{\bm{p},\sigma}c_{\bm{p},\sigma}
  -U\sum_{\bm{p},\bm{p}'\bm{q}}c^{\dagger}_{\bm{p}+\bm{q}/2,\uparrow}c^{\dagger}_{-\bm{p}+\bm{q}/2,\downarrow}
  c_{-\bm{p}'+\bm{q}/2,\downarrow}c_{\bm{p}'+\bm{q}/2,\uparrow}. \label{Hamiltonian}
\end{align}
Here, $c^{\dagger}_{\bm{p},\sigma}$ is a creation  operator of a Fermi
atom with  pseudospin $\sigma =\uparrow,\downarrow$ and the two-dimensional
momentum $\bm{p}=(p_{x},p_{y})$. $\xi_{\bm p}=p^2/(2m)-\mu$ is the
kinetic energy, measured from the Fermi chemical potential $\mu$, where
$m$ is an atomic mass. The pairing interaction $-U$ ($<0$) is assumed to
be tunable by adjusting the threshold energy of a Feshbach resonance. As
usual, we measure the interaction strength in terms of the
two-dimensional $s$-wave scattering length $a_{2{\rm D}}$, which is
related to $U$ as\cite{Morgan},
\begin{equation}
\frac{1}{U}=\frac{m}{2\pi}\ln{(k_{\rm F}a_{2{\rm D}})}+\sum_{p\geq
 k_{\rm F}}\frac{m}{p^2}, \label{U}
\end{equation}
where $k_{\rm F}=\sqrt{2\pi N}$ is
the Fermi momentum, with $N$ being the total number of Fermi atoms.
Using this scale, $\ln{(k_{\rm F}a_{2{\rm D}})} \ll
-1$ ($ \gg 1$)
corresponds to the strong-coupling (weak-coupling) regime.
$-1\lesssim \ln{(k_{\rm F}a_{2{\rm D}})} \lesssim 1$ is the crossover region.
\par
\begin{figure}[t]
\begin{center}
\includegraphics[width=1\textwidth]{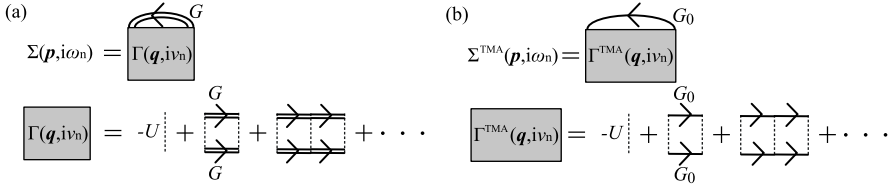}
\caption{
(a) Self-energy $\Sigma({\bm p},i\omega_n)$ in the self-consistent
 $T$-matrix approximation (SCTMA).
 The double-solid line shows the dressed Green's function $G$ in SCTMA.
 The dotted line represents the pairing interaction $-U$.
 We also show the self-energy $\Sigma^{\rm TMA}({\bm p},i\omega_n)$ in
 the non-self-consistent $T$-matrix approximation (TMA) in (b),
 where all $G$ 
 appearing in (a) are replaced
 by the free propagator $G_{0}=1/(i\omega_{n}-\xi_{\bm{p}})$.}
\label{diagram}       
\end{center}
\end{figure}
\par
Many-body corrections to Fermi single-particle excitations can be
conveniently incorporated into the theory by considering  the self-energy $\Sigma(\bm{p},i\omega_{n})$ in the single-particle thermal Green's function,
\begin{equation}
G(\bm{p},i\omega_{n})=
{1 \over i\omega_n-\xi_{\bm p}-\Sigma(\bm{p},i\omega_{n})}.
\label{Green}
\end{equation}
Here, $\omega_n$ is the fermion Matsubara frequency.  The self-energy
$\Sigma(\bm{p},i\omega_{n})$ in the
self-consistent $T$-matrix approximation (SCTMA) is diagrammatically described as
Fig. \ref{diagram}(a), which gives \cite{Haussmann,Bauer,Mulkerin},
\begin{equation}
\Sigma({\bm p},i\omega_n)
=T\sum_{\bm{q},i\nu_{n}}
\Gamma(\bm{q},i\nu_{n})G(\bm{q}-\bm{p},i\nu_{n}-i\omega_{n}).
\label{Sigma}
\end{equation}
Here, $\nu_{n}$ is the boson Matsubara frequency. 
The particle-particle scattering matrix $\Gamma(\bm{q},i\nu_{n})$ in
SCTMA has the form (see the second line in Fig.\ref{diagram}(a))
\begin{equation}
\Gamma(\bm{q},i\nu_{n})=-\frac{U}{1-U\Pi(\bm{q},i\nu_{n})}, 
\label{Gamma}
\end{equation}
where
\begin{equation}
\Pi(\bm{q},i\nu_{n})=T\sum_{\bm{p},i\omega_{n}}G\left(\bm{p}+\frac{\bm{q}}{2},i\nu_{n}+i\omega_{n}\right)G\left(-\bm{p}+\frac{\bm{q}}{2},-i\omega_{n}\right)
\label{PII}
\end{equation}
is a pair-correlation function, describing fluctuations in the Cooper
channel.
\par
The self-energy $\Sigma^{\rm TMA}(\bm{p},i\omega_{n})$ in the non-self-consistent $T$-matrix
approximation (TMA) is 
given by replacing all the 
 the dressed Green's functions in the SCTMA $\Sigma(\bm{p},i\omega_{n})$
by the
free Fermi Green's functions
$G_{0}(\bm{p},i\omega_{n})=(i\omega_{n}-\xi_{\bm{p}})^{-1}$, as shown in
Fig. \ref{diagram} (b).
That is, 
\begin{equation}
\Sigma^{\rm TMA}({\bm p},i\omega_n)
=T\sum_{\bm{q},i\nu_{n}}
\Gamma^{\rm
TMA}(\bm{q},i\nu_{n})G_0(\bm{q}-\bm{p},i\nu_{n}-i\omega_{n}),
\label{Sigma_TMA}
\end{equation}
where 
$
\Gamma^{\rm TMA}(\bm{q},i\nu_{n})=-U/(1-U\Pi^{\rm
TMA}(\bm{q},i\nu_{n}))
$
and the TMA pair correlation function is given by
$
\Pi^{\rm
TMA}(\bm{q},i\nu_{n})=T\sum_{\bm{p},i\omega_{n}}G_0\left(\bm{p}+\frac{\bm{q}}{2},i\nu_{n}+i\omega_{n}\right)G_0\left(-\bm{p}+\frac{\bm{q}}{2},-i\omega_{n}\right)
$.
Because of this simplification, in contrast to SCTMA,
strong coupling corrections to Fermi single-particle excitations, as
well as the resulting pseudogap phenomenon, are completely ignored in
evaluating the TMA particle-particle scattering matrix $\Gamma^{\rm
TMA}(\bm{q},i\nu_{n})$.
We will show how this ignorance leads to the overestimation of the
pseudogap phenomenon in the weak-coupling case when $\ln{(k_{\rm
F}a_{2{\rm D}})}\gtrsim 0$.
\par
In both SCTMA and TMA, the Fermi chemical potential $\mu$ is determined from
the equation of the total number $N$ of Fermi atoms,
\begin{equation}
N=2T\sum_{\bm{p},i\omega_{n}}G(\bm{p},i\omega_{n}). 
\label{Number}
\end{equation}
We then examine the pseudogap appearing in the single-particle 
density of states $\rho(\omega)$, given 
by 
\begin{equation}
 \rho(\omega)=-\frac{1}{\pi}\sum_{\bm{p}}{\rm
 Im}G(\bm{p},i\omega_{n}\rightarrow \omega +i\delta). \label{Dos}
 \end{equation}
\par
We briefly note that neither SCTMA nor TMA can describe the BKT phase
transition temperature $T_{\rm BKT}$.
Thus, this paper only deals with the normal phase above $T_{\rm BKT}$.
\par
\section{Pseudogap Phenomena in the crossover regime}
\par
\par
\par
\begin{figure}[t]
\begin{center}
\includegraphics[width=0.54\textwidth]{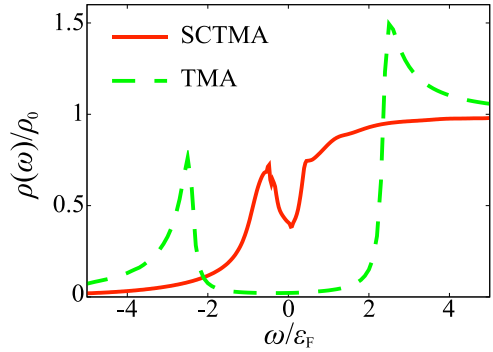}
\caption{ Calculated density of states $\rho(\omega)$ in a
 two-dimensional Fermi gas.
 The solid line and the dashed line show the results in SCTMA and TMA,
 respectively.
 $\rho_{0}=m/2\pi$ is the density of state in a two-dimensional free
 Fermi gas.
 We set $\ln{(k_{\rm F}a_{2{\rm D}})}=0.57$, and
 $T=T^{\rm exp}_{\rm BKT}=0.146T_{\rm F}$, where
 $T^{\rm exp}_{\rm BKT}$ is the observed BKT phase transition
 temperature at this interaction strength in a $^6$Li Fermi
 gas\cite{Ries,Murthy2}. (Color figure online.)}
\label{Dos_Tbkt}    
\end{center}
\end{figure}
\par
\par
Figure \ref{Dos_Tbkt} shows the density of states (DOS) $\rho(\omega)$
in a two-dimensional Fermi gas,
when 
$\ln{(k_{\rm F}a_{2{\rm D}})}=0.57$ (in the crossover region) at the
observed BKT phase transition temperature $T^{\rm exp}_{\rm
BKT}=0.146T_{\rm F}$ (where $T_{\rm F}$ is
 the Fermi temperature) in a $^6$Li Fermi gas\cite{Ries,Murthy2}.
We see that SCTMA gives a  pseudogap, that is, a dip structure
around $\omega=0$.
As discussed in the three-dimensional case \cite{Tsuchiya}, this dip
structure originates from pairing fluctuations around the Fermi surface,
and the resulting formation of preformed Cooper pairs.
Such an anomalous structure is also seen in the case of TMA, as shown in
Fig. \ref{Dos_Tbkt}.
However, the pseudogap structure in this case is much more remarkable
than that in the case of SCTMA, and the overall structure is rather
close to the BCS-type superfluid density of states 
with the coherence peaks of the gaps
edges (although the system in the present case is still in the normal
state).
At this interaction strength, the binding energy $E_{\rm b}=1/(ma_{\rm
2D}^2)$ 
of a two-body bound state equals $E_{\rm b}=0.64\varepsilon_{\rm F}$
(where $\varepsilon_{\rm F}$ is the Fermi energy).
While this value is comparable to the pseudogap size seen in
$\rho(\omega)$ in SCTMA in Fig. \ref{Dos_Tbkt}, the peak-to-peak energy
in $\rho(\omega)$ in TMA ($\gtrsim 4\varepsilon_{\rm F}$) is much larger
than $E_{\rm b}$.
This implies that the pseudogap size in TMA does not reflect the binding
energy of a preformed pair in this regime.
\par
\par
\begin{figure}[t]
\begin{center}
\includegraphics[width=0.95\textwidth]{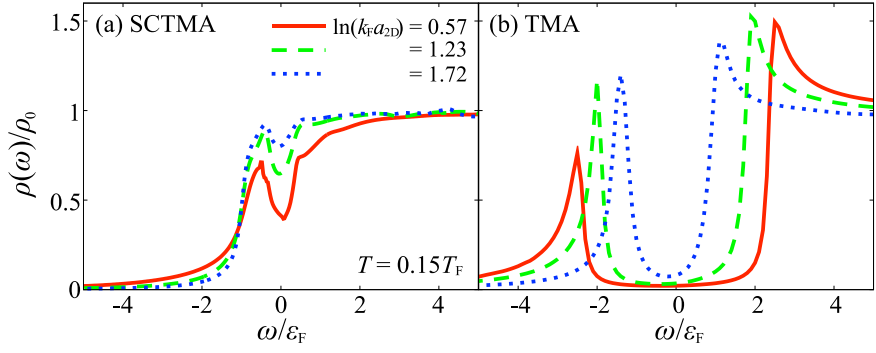}
\caption{Calculated density of states $\rho(\omega)$ in a
 two-dimensional Fermi gas at $T=0.15T_{\rm F}$  (a) SCTMA. (b) TMA.
(Color figure online.)
 }
\label{Dos_nearTbkt}       
\end{center}
\end{figure}

Figure \ref{Dos_nearTbkt} shows the interaction dependence of the
density of states $\rho(\omega)$
when  $T=0.15T_{\rm F}$.
In the case of SCTMA shown in panel (a), the dip structure becomes less
remarkable with decreasing the interaction strength, as
expected. 
According to the preformed pair scenario for the pseudogap phenomenon
\cite{Tsuchiya}, the pseudogap gradually disappears when $T\gtrsim
E_{\rm b}$.
Noting that
$E_{\rm b}(\ln{(k_{\rm F}a_{2{\rm D}})}=1.23)=0.17\varepsilon_{\rm F}$,
and $E_{\rm b}(\ln{(k_{\rm F}a_{2{\rm D}})}=1.72)=0.064\varepsilon_{\rm
F}$, one finds that the interaction dependence of the pseudogap
structure seen in Fig. \ref{Dos_nearTbkt} (a) is consistent with this
scenario. 

However, a large gap still remains
in the case of TMA 
even in the  weak coupling case when
$\ln{(k_{\rm  F}a_{2{\rm D}})}=1.72$, as shown in
Fig. \ref{Dos_nearTbkt} (b).
This is clearly contradict with the ordinary pseudogap case because
$T=0.15T_{\rm F}$ is much larger than the binding energy
$E_{\rm b}=0.064\varepsilon_{\rm F}$ at this interaction strength.
 \par
\par
\par

\begin{figure}[t]
\begin{center}
\includegraphics[width=0.54\textwidth]{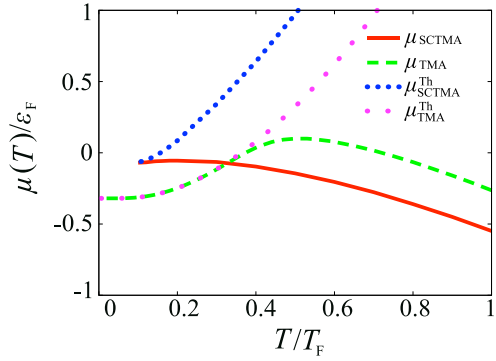}
\caption{Calculated chemical potential as a function of
 temperature, when $\ln{(k_{\rm F}a_{2{\rm D}})}=0.57$.
 $\mu_{\rm SCTMA}$ and $\mu_{\rm TMA}$ show the solution for the number
 equation (\ref{Number}) in SCTMA and TMA, respectively.
 $\mu^{\rm Th}_{\rm SCTMA} (\mu_{\rm TMA}^{\rm Th})$ satisfies the
 Thouless criterion in Eq. (\ref{Thouless}) in SCTMA (TMA). (Color
 figure online.) 
 }
\label{mu}    
\end{center}
\end{figure}

To explain the reason why TMA gives very different results form those in
SCTMA in the weak-coupling regime ($\ln{(k_{\rm F}a_{2{\rm D}})}\gtrsim
0$), it is instructive to consider 
the weak-coupling limit [$\ln{(k_{\rm F}a_{2{\rm D}})}\rightarrow \infty$] at $T=0$,
where the system should become a free Fermi gas with no pseudogap.
In a two-dimensional uniform system,
although the Hohenberg-Mermin-Wagner theorem\cite{Mermin,Hohenberg}
prohibits the long-range superfluid order at $T> 0$,
it may be realized at $T=0$, when the 
Thouless criterion \cite{Thouless},
\begin{equation}
\Gamma^{-1}(\bm{q}=\bm{0},i\nu_{n}=0)=0 \label{Thouless},
\end{equation}
is satisfied.
When one use $\Gamma^{\rm TMA}(\bm{q},i\nu_{n})$ given below Eq. (\ref{Sigma_TMA}), the
TMA Thouless criterion gives the chemical potential
$\mu_{\rm TMA}(T=0)=-E_{\rm b}/2$, indicating that all the
Fermi atoms form two-body bound molecules with the binding energy
$E_{\rm b}=1/ma_{\rm 2D}^2$.
Even not in the weak-coupling limit but at
$\ln{(k_{\rm F}a_{2{\rm D}})}=0.57$,
$\mu_{\rm TMA}(T)$ is found to approach $-E_{\rm b}/2\simeq
-0.32\varepsilon_{\rm F}$ at low temperatures, as seen in Fig. \ref{mu}.
 When the Thouless criterion in Eq. (\ref{Thouless}) is satisfied,
one may approximate the self-energy in Eq. (\ref{Sigma_TMA})
to $\Sigma^{\rm TMA}(\bm{p},i\omega_{n})\simeq -\Delta^2_{\rm
PG}G_{0}(-\bm{p},-i\omega_{n})$,
where 
$\Delta_{\rm PG}=\sqrt{-T\sum_{\bm{q},i\nu_{n}}\Gamma(\bm{q},i\nu_{n})}$
is sometimes referred to as the pseudogap parameter in the literature\cite{Tsuchiya,Chen,Matsumoto1}.
In this so-called static approximation,
the TMA Green's function is approximated to
\begin{equation}
G_{\rm
 TMA}(\bm{p},i\omega_{n})=-\frac{i\omega_{n}+\xi_{\bm{p}}}{\omega^2_{n}+\xi^2_{\bm{p}}+\Delta^2_{\rm
 PG}}. \label{Green_TMA}
\end{equation}
Equation (\ref{Green_TMA}) just has the same form as the diagonal
component of the mean-filed BCS Green's function \cite{Schrieffer}, so that one
has a clear energy gap with $E_{\rm PG}=2\sqrt{|\mu_{\rm TMA}|^2+\Delta_{\rm
PG}^2}$.
In addition, substituting Eq. (\ref{Green_TMA}) into the number equation
(\ref{Number}) at $T=0$, one obtains $\Delta_{\rm PG}=2\sqrt{\varepsilon_{\rm F}(\varepsilon_{\rm
F}-\mu_{\rm TMA})}$, unphysically giving the large (pseudo) gap size as
$E_{\rm PG}=4\varepsilon_{\rm F}\gg 2E_{\rm b}$, even in the
weak-coupling limit.

In the case of SCTMA, the static approximation for the SCTMA Green's function
gives, 
\begin{equation}
G_{\rm
 SCTMA}(\bm{p},i\omega_{n})=-\frac{i\omega_{n}+\xi_{\bm{p}}}{\omega^2_{n}+\xi^2_{\bm{p}}+2\Delta^2_{\rm
 PG}\left[1+\sqrt{1+\frac{4\Delta^2_{\rm
 PG}}{\omega^2_{n}+\xi^2_{\bm{p}}}}\right]^{-1}  }. \label{Green_SCTMA}
\end{equation}
Apart from the factor $2\left[1+\sqrt{1+\frac{4\Delta^2_{\rm
 PG}}{\omega^2_{n}+\xi^2_{\bm{p}}}}\right]^{-1}$,
 Eq. (\ref{Green_SCTMA}) is still close to the diagonal component of the
 mean-filed BCS Green's function.
 Indeed, when one uses Eq. (\ref{Green_SCTMA}) to evaluate the number
 equation (\ref{Number}), together with the Thouless criterion in
 Eq. (\ref{Thouless}), the resulting coupled equations are found to be
 formally close to the number equation at the gap equation in the
 mean-field BCS theory at $T=0$\cite{Miyake}, giving $\mu_{\rm
 SCTMA}=\varepsilon_{\rm F}$, and $\Delta_{\rm PG}=0$, as expected.
\par
\par
\par
\par
 The above discussions at $T=0$ may be also applicable to the
 weak-coupling regime at the finite temperatures.
 In this case, although the Thouless criterion is, exactly speaking, not
 satisfied, the TMA chemical potential $\mu_{\rm TMA}$ becomes very
 close to the value $\mu^{\rm Th}_{\rm TMA}$, (which satisfies
 Eq. (\ref{Thouless}) (where $\Gamma^{\rm TMA}$ is used).) at low
 temperatures, as shown in Fig. \ref{mu}.
 In the case of Fig. \ref{mu}, the static approximation is considered
 to be valid for $T\lesssim 0.4T_{\rm F}$, where an unphysically large
 pseudogap is expected in TMA density of states.
 In the case of SCTMA, $\mu_{\rm SCTMA}$ becomes close to the value
 $\mu^{\rm Th}_{\rm SCTMA}$, which satisfies the Thouless criterion in
 Eq (\ref{Thouless}) (where $\Gamma$ in SCTMA is used),
 only when $T/T_{\rm F}\lesssim 0.1$, so that the pseudogap structure in
 this case is not so remarkable as that in the TMA case.

 Physically, when the Fermi chemical potential approximately satisfies
 the Thouless criterion $(\mu \simeq \mu^{\rm Th})$,
 strong pairing fluctuations cause a dip structure in the density of
 states $\rho(\omega)$ around $\omega=0$.
 However, in the weak-coupling regime, since preformed pairs are
 dominantly formed around the Fermi surface, the appearance of the
 pseudogap would also suppress pairing fluctuations, as well as the
 pseudogap phenomenon.
 Such a feedback effect is, however, completely ignored in TMA, because
 the free propagator $G_{0}$ with no TMA self-energy is used in
 evaluating the particle-particle scattering matrix
 $\Gamma(\bm{q},i\nu_{n})$.
 In this case, SCTMA treats pairing fluctuations in a consistent
 manner, so that the expected pseudogap behavior of the density of
 states is correctly obtained in the weak-coupling regime.
\par
\section{Summary}
\par
To summarize, we have discussed the pseudogap phenomenon in a
two-dimensional ultracold Fermi gas in the crossover region, as well as
in the weak-coupling regime.
We showed that the pseudogap phenomenon associated with pairing
fluctuations in this regime can correctly be treated by the
self-consistent $T$-matrix approximation (SCTMA).
In contrast to the ordinary (non-self-consistent) $T$-matrix
approximation (TMA), which unphysically gives a large pseudogap in
the density of states even in the weak-coupling regime,
SCTMA gives a expected small pseudogap, which gradually
disappears as one approaches the weak-coupling regime.
 We also pointed out the importance of a feedback effect in theoretically
 dealing with pairing fluctuations in this regime, which is completely
 ignored in TMA.
\par
\par
\begin{acknowledgements}
We thank H. Tajima,  T. Yamaguchi, P. van Wyk , and D. Kagamihara for discussions.
 M. M. was supported by  Graduate School Doctoral Student Aid Program
 from Keio University.
 R. H. was supported by a Grant-in-Aid for JSPS fellows.
 D. I. was supported by Grant-in-Aid for Young Scientists (B)
 (No.16K17773) from JSPS in Japan.
 This work was supported by the KiPAS project in Keio university. Y.O
 was supported by Grant-in-Aid for Scientific Research from MEXT and
 JSPS in Japan (No.15K00178, No.15H00840, No.16K05503).
\end{acknowledgements}
\par
\par


\begin{thebibliography}{99}
\bibitem{Gurarie} V.  Gurarie, and L. Radzihovsky, Ann. Phys. {\bf 332}, 2 (2007).
\bibitem{Bloch} I. Bloch, J. Dalibard, and W. Zwerger, Rev. Mod. Phys. {\bf 80}, 885 (2008).	 
\bibitem{Stewart}J.T. Stewart, J.P. Gaebler, D.S. Jin, Nature {\bf 454},
	744 (2008)
\bibitem{Gaebler} J.P. Gaebler, {\it et al}., Nat. Phys. {\bf 6}, 569
	(2010).	
\bibitem{Martiyanov} K. Martiyanov, V. Makhalov, and A. Turlapov,
	Phys. Rev. Lett. {\bf 105}, 030404 (2010).
\bibitem{Feld} M. Feld, {\it et al}., Nature \textbf{480}, 75 (2011).	
\bibitem{Frohlich} B. Fr\"{o}hlich, {\it et al}., Phys. Rev. Lett. {\bf 106}, 105301 (2011).		
 \bibitem{Sommer} A. T. Sommer, {\it et al}., Phys. Rev. Lett. {\bf 108}, 045302 (2012).
\bibitem{Makhalov} V. Makhalov, K. Martiyanov, and A. Turlapov,
	Phys. Rev. Lett. {\bf 112}, 045301 (2014).
\bibitem{Ries} M. G. Ries, {\it et. al.}, Phys. Rev. Lett. {\bf 114}, 230401 (2015).
 \bibitem{Murthy2} P. A. Murthy, {\it et al.}, Phys. Rev. Lett. {\bf
	 115}, 010401 (2015).
\bibitem{Fenech} K. Fenech, {\it et al}., Phys. Rev. Lett. {\bf 116}, 045302
	(2016).
\bibitem{Berezinskii} V. L. Berezinskii, Sov. Phys. JETP {\bf 32}, 493 (1971).
\bibitem{Kosterlitz} J. M. Kosterlitz, and D. J. Thouless, J. Phys. C
	{\bf 6}, 1181 (1973).
\bibitem{Tsuchiya} S. Tsuchiya, R. Watanabe, and Y. Ohashi, Phys. Rev. A
	{\bf 80}, 033613 (2009).
\bibitem{Chen} Q. J. Chen, and K. Levin, Phys. Rev. Lett. {\bf 102},
	 190402 (2009).	
 \bibitem{HuiHu} H. Hu, X. -J. Liu, P. D. Drummond, and H. Dong, Phys. Rev. Lett. {\bf 104}, 240407 (2010).
\bibitem{Marsiglio} F. Marsiglio, {\it et al.}, Phys. Rev. B {\bf 91},	054509 (2015).
 \bibitem{Matsumoto1}  M. Matsumoto, D. Inotani, and Y. Ohashi,
	Phys. Rev. A {\bf 93}, 013619 (2016).			
\bibitem{Morgan} S. A. Morgan, M. D. Lee, and K. Burnett, Phys. Rev. A
	{\bf 65}, 022706 (2002).
\bibitem{Haussmann} R. Haussmann, Z. Phys. B: Condens. Matter {\bf 91},
	291 (1993).
\bibitem{Bauer} M. Bauer, M. M. Parish, T. Enss, Phys. Rev. Lett. {\bf
	112}, 135302 (2014).
\bibitem{Mulkerin} B. C. Mulkerin, {\it et al.}, Phys. Rev. A {\bf 92}, 063636 (2015).	
\bibitem{Mermin} N. D. Mermin, and H. Wagner, Phys. Rev. Lett. {\bf 17}, 1133 (1966).
\bibitem{Hohenberg} P. C. Hohenberg, Phys. Rev. {\bf 158}, 383 (1967).
\bibitem{Thouless} D. J. Thouless, Ann. Phys. {\bf 10}, 553 (1960).	
\bibitem{Schrieffer} J. R. Schrieffer, {\it Theory of Superconductivity} (Addison-Wesley, NY, 1964).		 
\bibitem{Miyake}  K. Miyake, Prog. Theor. Phys. {\bf 69}, 6 (1983).	
\end{thebibliography}
\end{document}